\documentclass[%
 reprint,
showpacs,preprintnumbers,
 amsmath,amssymb,
onecolumn]{revtex4}

\usepackage{slashed}
\usepackage{graphicx}
\usepackage{dcolumn}
\usepackage{bm}

\begin{document}
\title{Locating the critical end point using the linear sigma model coupled to quarks.}
%
%

\author{Alejandro Ayala$^{1,3}$, Jos\'e Antonio Flores$^2$, Luis Alberto Hern\'andez$^1$ and Sa\'ul Hern\'andez-Ortiz$^1$}
\affiliation{
$^1$Instituto de Ciencias
Nucleares, Universidad Nacional Aut\'onoma de M\'exico, Apartado
Postal 70-543, M\'exico Distrito Federal 04510,
Mexico.\\
$^2$Departamento de F\'isica, Facultad de Ciencias, Avenida Universidad 3000, Universidad Nacional Aut\'onoma de M\'exico, Ciudad Universitaria, 04510 Distrito Federal, M\'exico\\
$^3$Centre for Theoretical and Mathematical Physics, and Department of Physics,
University of Cape Town, Rondebosch 7700, South Africa}

\begin{abstract}%
We use the linear sigma model coupled to quarks to compute the effective potential beyond the mean field approximation, including the contribution of the ring diagrams at finite temperature and baryon density. We determine the model couplings and use them to study the phase diagram in the baryon chemical potential-temperature plane and to locate the Critical End Point.
\end{abstract}

\maketitle
\section{Introduction}
The study of strongly interacting matter under extreme conditions, such as high temperature ($T$) and baryon chemical potential ($\mu_B$), is of great importance in today's physics. One of the principal goals is to gather accurate knowledge of the phase diagram in the $\mu-T$ plane. In this work we use the linear sigma model coupled to quarks, including the plasma screening effects, to explore the effective QCD phase diagram, focusing on the chiral symmetry restoration.  To do so, we fix the coupling constants using the physical values of the model parameters, such as the vacuum pion and sigma masses, the critical temperature $T_c$ at $\mu_B$ = 0 and the conjectured maximum value of $\mu_B$ of the transition line at $T = 0$.
\label{intro}

\section{\label{sec:LSMq}Linear Sigma Model coupled to quarks}
In order to study the spontaneous breaking of chiral symmetry with the intention of sketching the QCD phase diagram at finite temperature and quark chemical potential. We use the \textit{Linear Sigma Model} coupled to quarks, that is an effective model that accounts for the physics of spontaneous symmetry breaking. The Lagrangian for the linear sigma model when the two lightest quarks are included is given by
\begin{align}
   \mathcal{L}\ =&\ \frac{1}{2}(\partial_\mu \sigma)^2  + \frac{1}{2}(\partial_\mu \vec{\pi})^2 + \frac{a^2}{2} (\sigma^2 + \vec{\pi}^2) - \frac{\lambda}{4} (\sigma^2 + \vec{\pi}^2)^2 \nonumber \\
   &\ + i \bar{\psi} \gamma^\mu \partial_\mu \psi -g\bar{\psi} (\sigma + i \gamma_5 \vec{\tau} \cdot \vec{\pi} )\psi ,
\label{lagrangian}
\end{align}
where $\psi$ is an SU(2) isospin doublet, $\vec{\pi}=(\pi_1, \pi_2, \pi_3 )$ is an isospin triplet and $\sigma$ is an isospin singlet. $\lambda$ is the boson's self-coupling and $g$ is the fermion-boson coupling. $a^2>0$ is the mass parameter.  To allow for an spontaneous breaking of symmetry, we let the $\sigma$ field to develop a vacuum expectation value $v$
\begin{equation}
   \sigma \rightarrow \sigma + v,
\label{shift}
\end{equation}
which can later be taken as the order parameter of the theory.  After this shift, the Lagrangian can be rewritten as
\begin{align}
   {\mathcal{L}}\ = &\ -\frac{1}{2}[\sigma \partial_{\mu}^{2}\sigma]-\frac{1}{2}\left(3\lambda v^{2}-a^{2} \right)\sigma^{2}\nonumber \\
   &\ -\frac{1}{2}[\vec{\pi}\partial_{\mu}^{2}\vec{\pi}]-\frac{1}{2}\left(\lambda v^{2}- a^2 \right)\vec{\pi}^{2}+\frac{a^{2}}{2}v^{2}\nonumber \\
  &\ -\frac{\lambda}{4}v^{4} + i \bar{\psi} \gamma^\mu \partial_\mu \psi -gv \bar{\psi}\psi + {\mathcal{L}}_{I}^b + {\mathcal{L}}_{I}^f,
  \label{lagranreal}
\end{align}
where ${\mathcal{L}}_{I}^b$ and  ${\mathcal{L}}_{I}^f$ are given by
\begin{align}
  {\mathcal{L}}_{I}^b\ = &\ -\frac{\lambda}{4}\Big[(\sigma^2 + (\pi^0)^2)^2 + 4\pi^+\pi^-(\sigma^2 + (\pi^0)^2 + \pi^+\pi^-)\Big],\nonumber \\
  {\mathcal{L}}_{I}^f\ =&\ -g\bar{\psi} (\sigma + i \gamma_5 \vec{\tau} \cdot \vec{\pi} )\psi.
  \label{lagranint}
\end{align}

Equation~(\ref{lagranint}) describes the interactions among the $\sigma$, $\vec{\pi}$ and $\psi$ fields after symmetry breaking. From Eq.~(\ref{lagranreal}) one can see that the sigma, the three pions and the quarks have masses given by
\begin{equation}
  m^{2}_{\sigma}\ =\ 3  \lambda v^{2}-a^{2},\quad  m^{2}_{\pi}\ =\ \lambda v^{2}-a^{2},\quad  m_{f}\ =\ gv,
\label{masses}
\end{equation}
respectively. We study the behavior of the effective potential, which we deduce in the next section in detail, in order to analyze the chiral symmetry restoration conditions in terms of temperature and quark chemical potential.

\section{\label{EffPot}Effective potential}
In this section, we compute the $T-$ and $\mu-$dependent effective potential up to ring diagrams in order to account for the plasma screening effects. The tree level potential is given by

\begin{equation}
    V^{\text{tree}}(v)=-\frac{a^2}{2}v^2+\frac{\lambda}{4}v^4,
    \label{treelevel}
\end{equation}
whose minimum is given by

\begin{equation}
    v_0=\sqrt{\frac{a^2}{\lambda}},
\end{equation}
since $v_0\neq 0$, we notice that the symmetry is spontaneously broken. To include quantum corrections at finite temperature and density, we work within the imaginary-time formalism of thermal field theory. The general expression for the one-loop boson contribution can be written as

\begin{equation}
    V^{(1)\text{b}}(v,T)=T\sum_n\int\frac{d^3k}{(2\pi)^3} \ln D(\omega_n,\vec{k})^{1/2},
    \label{1loopboson}
\end{equation}
where

\begin{equation}
D(\omega_n,\vec{k})=\frac{1}{\omega_n^2+k^2+m_b^2},
\end{equation} 
is the free boson propagator with $m_b^2$ being the square of the boson's mass and $\omega_n=2n\pi T$ the Matsubara frequencies for boson fields. 

For a fermion field with mass $m_f$, the general expression for the one-loop correction at finite temperature and quark chemical potential $\mu_q$ is

\begin{equation}
    V^{(1)\text{f}}(v,T,\mu_q)=-T\sum_n\int\frac{d^3k}{(2\pi)^3} \text{Tr}[\ln S(\tilde{\omega}_n-i\mu_q,\vec{k})^{-1}],
    \label{1loopfermion}
\end{equation}
where

\begin{equation}
S(\tilde{\omega}_n-i\mu_q,\vec{k})=\frac{1}{\gamma_0 \left (\tilde{\omega}_n-i\mu_q\right )+\not{\!k}+m_f},
\end{equation}
is the free fermion propagator and $\tilde{\omega}_n=(2n+1)\pi T$ are the Matsubara frequencies for fermion fields. The ring diagrams term is given by

\begin{equation}
    V^{\text{Ring}}(v,T,\mu_q)=\frac{T}{2}\sum_n\int\frac{d^3k}{(2\pi)^3}\ln (1+\Pi(m_b,T,\mu_q)D(\omega_n,\vec{k})),
    \label{rings}
\end{equation}
where $\Pi(m_b,T,\mu_q)$ is the boson's self-energy. In order to compute the self-energy for one boson field, we include all the contribution from the Feynman rules. Therefore, the self-energy is written as

\begin{equation}
    \Pi(T,\mu_q)=\sum_{i=\sigma, \pi^0,\pi^\pm}\Pi_{\text{i}}(T)+\sum_{j=u,d}\Pi_{\text{j}}(T,\mu_q),
\end{equation}
where
\begin{eqnarray}	
    \Pi_\sigma(T)&=&\frac{\lambda}{4}\left [12 I\left (\sqrt{m_\sigma^2+\Pi_\sigma}\right )+4I\left (\sqrt{m^2_{\pi^0}+\Pi_{\pi^0}}\right ) +8I\left (\sqrt{m^2_{\pi^\pm}+\Pi_{\pi^\pm}}\right )\right ],\nonumber \\
&&\nonumber \\
    \Pi_{\pi^\pm}(T)&=&\frac{\lambda}{4}\left [4 I\left (\sqrt{m_\sigma^2+\Pi_\sigma}\right )+4I\left (\sqrt{m^2_{\pi^0}+\Pi_{\pi^0}}\right ) +16I\left (\sqrt{m^2_{\pi^\pm}+\Pi_{\pi^\pm}}\right )\right ],\nonumber \\
&&\nonumber \\
    \Pi_{\pi^0}(T)&=&\frac{\lambda}{4}\left [4 I\left (\sqrt{m_\sigma^2+\Pi_\sigma}\right )+12I\left (\sqrt{m^2_{\pi^0}+\Pi_{\pi^0}}\right ) + 8I\left (\sqrt{m^2_{\pi^\pm}+\Pi_{\pi^\pm}}\right )\right ],
\end{eqnarray}
with
\begin{equation}
    I(x)=\frac{1}{2\pi^2}\int dk \frac{k^2}{\sqrt{k^2+x}} n\left (\sqrt{k^2+x}\right ),
\end{equation}
and $n(x)$ being the Bose-Einstein distribution. On the other hand, the fermion contribution is given by

\begin{align}
    \Pi_j(T,\mu_q)&=-g^2 T\sum_n\int \frac{d^3k}{(2\pi)^3}\text{Tr}[S(\tilde{\omega}_n-i\mu_q,\vec{k},m_f)\nonumber \\
    &\times S(\tilde{\omega}_n-i\mu_q-\tilde{\omega}_m,\vec{k}-\vec{p},m_f)].
    \label{selfEF}
\end{align}

As we work close to the phase transition, a good approximation is to take the fermion's mass $(m_f=0)$ and the boson's mass including thermal correction $(m_i^2+\Pi_i=0)$ to be small. Therefore, the total self-energy for one boson is

\begin{equation}
    \Pi(T,\mu_q)=-N_fN_cg^2\frac{T^2}{\pi^2}[\text{Li}_2(-e^{\mu_q/T})+\text{Li}_2(-e^{-\mu_q/T})]+\frac{\lambda T^2}{2}.
    \label{fullselfenergy}
\end{equation}

With the boson self-energy at hand we can study the properties of the effective potential. In order to work with analytical expressions we turn to study two cases: first the high temperature approximation $(T\gg m_b, \ \mu_q)$ and then the low temperature approximation $(T\ll m_b, \ \mu_q)$. In the following we compute explicitly both cases.

\subsection{\label{HTapprox}High temperature approximation}

As long as $T$ is the largest of the energy scales, a high temperature approximation is suited to study chiral symmetry restoration. Let's start from Eq.~(\ref{1loopboson}), the one-loop correction for boson fields. First we compute the sum over Matsubara frequencies

\begin{equation}
    V^{(1)\text{b}}(v,T)=\frac{1}{2\pi^2}\int dk \ k^2\Big \{ \frac{\sqrt{k^2+m_b^2}}{2} +T\ln \Big(1-e^{-\sqrt{k^2+m_b^2}/T} \Big) \Big \}.
    \label{V1bwosum}
\end{equation}

Notice that Eq.~(\ref{V1bwosum}) has two pieces, the first one is the \textit{vacuum} contribution and the second one is the \textit{matter} contribution, namely, the $T$-dependent correction. In order to compute the vacuum term, we regularize and renormalize the former employing dimensional regularization and the Minimal Subtraction scheme (MS), with the renormalization scale $\tilde{\mu}=e^{-1/2}a$. For the matter term, we take the approximation $m_b/T\ll 1$ and we include only the most dominant terms~\cite{Ayala1,Ayala2}. Taking all of this into account, the one-loop contribution to the effective potential from fermion fields is given by

\begin{align}
    V_{\text{HT}}^{(1)\text{b}}(v,T)=&-\frac{m_b^4}{64\pi^2}\Big[ \ln \Big( \frac{4\pi a^2}{m_b^2}\Big)-\gamma_E+\frac{1}{2} \Big]\nonumber \\
    &-\frac{m_b^4}{64\pi^2}\ln \Big( \frac{m_b^2}{(4\pi T)^2}\Big)-\frac{\pi^2 T^4}{90}\nonumber \\
    &+\frac{m_b^2 T^2}{24}-\frac{m_b^3 T}{12\pi}.
    \label{final1loopb}
\end{align}

For the case of the fermion one-loop contribution, we follow the procedure outlined for the boson case. For the matter term, we compute the integral in momentum taking into account the approximation where $m_f/T\ll 1$ and $\mu_q/T<1$, and we consider only the dominant terms. After computing the momentum integral we get

\begin{align}
    V_{\text{HT}}^{(1)\text{f}}(v,T)&=\frac{m_f^4}{16\pi^2}\Big[ \ln \Big( \frac{4\pi a^2}{m_f^2}\Big)-\gamma_E+\frac{1}{2} \Big]\nonumber \\
    &+\frac{m_f^4}{16\pi^2}\Big[\ln \Big( \frac{m_f^2}{(4\pi T)^2}\Big)
    -\psi^0\Big( \frac{1}{2}+\frac{\text{i}\mu}{2\pi T} \Big)-\psi^0\Big( \frac{1}{2}-\frac{\text{i}\mu}{2\pi T} \Big)\Big]\nonumber \\
    &-8m_f^2T^2\Big[ \text{Li}_2(-e^{\mu_q/T}) +\text{Li}_2(-e^{-\mu_q/T}) \Big]\nonumber \\
    &+32T^4\Big[ \text{Li}_4(-e^{\mu_q/T}) +\text{Li}_4(-e^{-\mu_q/T}) \Big].
    \label{final1loopf}
\end{align}

For the purposes of considering the plasma screening effects, we go beyond the mean field approximation. These can be accounted for by means of the ring diagrams. Since we are working in the high temperature approximation, we notice that the lowest Matsubara mode is the most dominant term~\cite{LeBellac}. Therefore, we do not need to compute the other modes and Eq.~(\ref{rings}) becomes

\begin{align}
    V^{\text{Ring}}(v,T,\mu_q)&=\frac{T}{2}\int\frac{d^3k}{(2\pi)^3}\ln (1+\Pi(T,\mu_q)D(\vec{k}))\nonumber \\
    &=\frac{T}{4\pi^2}\int dk \ k^2 \Big \{ \ln(k^2+m_b^2+\Pi(T,\mu_q)) -\ln(k^2+m_b^2) \Big\}.
    \label{rings2}    
\end{align}
From Eq.~(\ref{rings2}), we see that both integrands are almost the same except that one is modified by the self-energy and the other one is not. Thus, after integration, we obtain that the ring diagrams contribution is

\begin{equation}
    V^{\text{Ring}}(v,T,\mu_q)=\frac{T}{12\pi}(m_b^3-(m_b^2+\Pi(T,\mu_q))^{3/2}).
    \label{finalrings}
\end{equation}

With these pieces at hand, we can write the effective potential up to the ring diagrams contribution in the high temperature approximation. The effective potential in the high temperature approximation is given by

\begin{align}
    V_{\text{HT}}^{\text{eff}}(v,T,\mu_q)&=-\frac{(a^2+\delta a^2)}{2}v^2
    +\frac{(\lambda+\delta \lambda)}{4}v^4\nonumber \\
    &+\sum_{b=\sigma,\bar{\pi}}\Big\{-\frac{m_b^4}{64\pi^2}\Big[ \ln \Big( \frac{ a^2}{4\pi T^2}\Big)-\gamma_E+\frac{1}{2} \Big]\nonumber \\
    &-\frac{\pi^2 T^4}{90}+\frac{m_b^2 T^2}{24}-\frac{(m_b^2+\Pi(T,\mu_q))^{3/2} T}{12\pi}\Big\}\nonumber \\
    &+\sum_{f=u,d}\Big\{\frac{m_f^4}{16\pi^2}\Big[ \ln \Big( \frac{ a^2}{4\pi T^2}\Big)-\gamma_E+\frac{1}{2}\nonumber \\
    &-\psi^0\Big( \frac{1}{2}+\frac{\text{i}\mu_q}{2\pi T} \Big)-\psi^0\Big( \frac{1}{2}-\frac{\text{i}\mu_q}{2\pi T} \Big)\Big]\nonumber \\
    &-8m_f^2T^2\Big[ \text{Li}_2(-e^{\mu_q/T})+\text{Li}_2(-e^{-\mu_q/T}) \Big]\nonumber \\
    &+32T^4\Big[ \text{Li}_4(-e^{\mu_q/T})+\text{Li}_4(-e^{-\mu_q/T}) \Big]\Big\}.
    \label{finalHTpotential}
\end{align}

Notice that the potentially dangerous pieces coming from linear or cubic powers of the boson mass, that could become imaginary for certain values of $v$, are removed or replaced by the contribution of the ring diagrams~\cite{DJ}.

\subsection{\label{sec:level6}Low temperature approximation}

To have access to the region in the QCD phase diagram where $\mu_B$ is large and $T$ is small, we compute the effective potential in the approximation where $T$ is the soft scale in the system. We call this the low temperature approximation. For boson fields case, we include a boson chemical potential. We relate this to the conservation of an average number of particles and not to a conserved charge. Therefore, the one-loop contribution for boson fields after the sum over Matsubara frequencies is

\begin{align}
    V_{\text{LT}}^{(1)\text{b}}(v,T,\mu_b)&=\frac{1}{2\pi^2}\int dk \ k^2\Big \{ \sqrt{k^2+m_b^2}\nonumber \\
    &+T\ln \Big(1-e^{-(\sqrt{k^2+m_b^2}-\mu_b)/T} \Big)\nonumber \\
    &+T\ln \Big(1-e^{-(\sqrt{k^2+m_b^2}+\mu_b)/T} \Big) \Big \}.
    \label{V1bwosumLT}
\end{align}
In Eq.~(\ref{V1bwosumLT}), the matter contribution has two terms corresponding to particles and anti-particles. In this work, we follow the procedure used in Ref.~\cite{chilenos}. The general idea consists on developing a Taylor series around $T=0$ of the following expression

\begin{equation}
    V_{\text{LT}}^{(1)\text{b}}(v,T,\mu_b)=\int_{\frac{\mu_b-m_b}{T}}^\infty V_0^{\text{b}}(v,\mu_b+xT)h_B(x)dx,
    \label{1lLT}
\end{equation}
where $h_B(x)$ is the first derivative of the Bose-Einstein distribution and $V_0^{\text{b}}(v,\mu_b+xT)$ is the one-loop boson contribution evaluated at $T=0$, which is
\begin{align}
    V_{0}^{(1)\text{b}}(v,\mu_b)&=-\frac{m_b^4}{64\pi^2}\Big[ \ln\Big( \frac{4\pi a^2}{(\mu_b+\sqrt{\mu_b^2-m_b^2})^2} \Big)\nonumber \\
    &-\gamma_E+\frac{1}{2}\Big]+\frac{\mu_b\sqrt{\mu_b^2-m_b^2}}{96\pi^2}(2\mu_b^2-5m_b^2).
    \label{V1bLT}
\end{align}

From here, the Taylor series give us an expression of one-loop matter contribution from one boson field in the low temperature approximation

\begin{equation}
    V_{\text{LT}}^{(1)\text{b}}(v,T,\mu_b)=V_0^{\text{b}}(v,\mu_b)+\frac{\pi^2 T^2}{12}\frac{\partial^2}{\partial T^2}V_0^{\text{b}}(v,\mu_b) +\frac{7\pi^4 T^4}{1260}\frac{\partial^4}{\partial T^4}V_0^{\text{b}}(v,\mu_b).
    \label{1loopBLT}
\end{equation}
For fermion fields, we implement the low temperature approximation in the same way as we did for boson fields. We now develop a Taylor series around $T=0$ and we get the one-loop contribution for one fermion field in the low temperature approximation
 
\begin{equation}
    V_{\text{LT}}^{(1)\text{f}}(v,T,\mu_q)=V_0^{\text{f}}(v,\mu_q)+\frac{\pi^2 T^2}{6}\frac{\partial^2}{\partial T^2}V_0^{\text{f}}(v,\mu_q)+\frac{\pi^4 T^4}{360}\frac{\partial^4}{\partial T^4}V_0^{\text{f}}(v,\mu_q).
    \label{1loopFLT}
\end{equation}

Equations~(\ref{treelevel}),~(\ref{1loopBLT}) and~(\ref{1loopFLT}) provide the full expression for the effective potential in the low temperature approximation, which is given by

\begin{align}
    V_{\text{LT}}^{\text{eff}}(v,T,\mu_q,\mu_b)&=-\frac{(a^2+\delta a^2)}{2}v^2
    +\frac{(\lambda+\delta \lambda)}{4}v^4 \nonumber \\
    &-\sum_{i=\sigma,\bar{\pi}}\Big \{ \frac{m_i^4}{64\pi^2}\Big[ \ln \Big( \frac{4\pi^2 a^2}{(\mu_b+\sqrt{\mu_b^2-m_i^2})^2}\Big) -\gamma_E+\frac{1}{2}  \Big]\nonumber\\
    &-\frac{\mu_b\sqrt{\mu_b^2-m_i^2}}{24\pi^2}(2\mu_b^2-5m_i^2) -\frac{T^2\mu_b}{12}\sqrt{2\mu_b^2-5m_i^2}\nonumber \\
    &-\frac{\pi^2T^4\mu_b}{180}\frac{(2\mu_b^2-3m_i^2)}{(\mu_b^2-m_i^2)^{3/2}} \Big\}+N_c\sum_{f=u,d}\Big\{ \frac{m_f^4}{16\pi^2}\Big[ \ln \Big( \frac{4\pi^2 a^2}{(\mu_q+\sqrt{\mu_q^2-m_f^2})^2}\Big)\nonumber \\
    &-\gamma_E+\frac{1}{2}  \Big]-\frac{\mu_q\sqrt{\mu_q^2-m_f^2}}{24\pi^2}(2\mu_q^2-5m_f^2) -\frac{T^2\mu_q}{6}\sqrt{\mu_q^2-m_f^2}\nonumber \\
    &-\frac{7\pi^2T^4\mu_q}{360}\frac{(2\mu_q^2-3m_f^2)}{(\mu_q^2-m_f^2)^{3/2}} \Big\} 
    \label{VeffLT}
\end{align}

We are now able to analyze the QCD phase transition in the regions of the QCD phase diagram where the temperature is larger than the quark chemical potential and where the temperature is smaller than the quark chemical potential. To do so, first we need to determine the value of all the parameters involved in the linear sigma model under appropriate conditions. In the following section we proceed in this direction to determine the values of those parameters and in particular of the couplings $\lambda$ and $g$.

\section{\label{sec:level7}Coupling Constants}

The effective potential has tree free parameters which should be fixed: the two coupling constants $\lambda$ and $g$ and the square mass parameter $a^2$. In order to determine the square mass parameter, we use that the vacuum boson masses in Eq.~(\ref{masses}) are related by the expression

\begin{equation}
    a=\sqrt{\frac{m_\sigma^2-3m_\pi^2}{2}}.
    \label{afixed}
\end{equation}

Therefore, we can fix $a$ by using the vacuum sigma and pion masses. To fix the coupling constants we use physical inputs from QCD matter around the phase transition in the high and low temperature domains.

\begin{figure}[t]
\begin{center}
\includegraphics[scale=0.45]{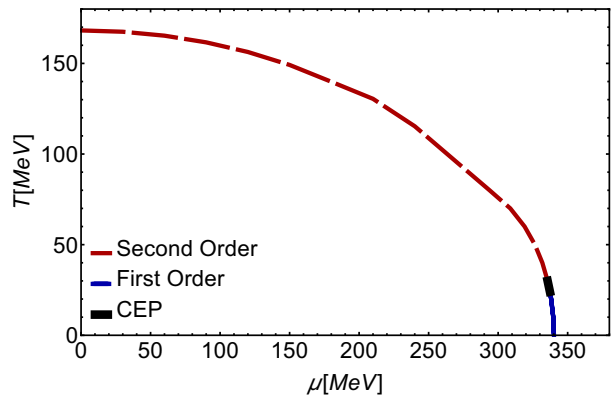}
\end{center}
\caption{QCD phase diagram obtained from the solutions to the equations that determine the coupling constants. This are computed with $T^c_0(\mu_q=0)=170$ MeV and $\mu^c_q(T=0)=340$ MeV. The second order transitions are indicated by red line and the first order transitions by the blue line. These areas represent the results directly obtained from our analysis.}
\label{FPD1}
\end{figure}
\begin{figure}[t]
\begin{center}
\includegraphics[scale=0.45]{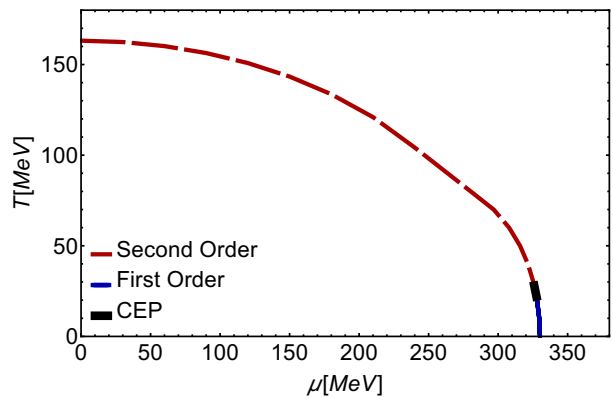}
\end{center}
\caption{QCD phase diagram obtained from the solutions to the equations that determine the coupling constants. This are computed with $T^c_0(\mu_q=0)=165$ MeV and $\mu^c_q(T=0)=330$ MeV. The second order transitions are indicated by red line and the first order transitions by the blue line. These areas represent the results directly obtained from our analysis.}
\label{FPD2}
\end{figure}

First from LQCD computations~\cite{latticeTc}, we know that at $\mu_q\equiv\mu_B/3=0$, the QCD phase transition is a crossover (in our case a second order transition), and happens for the case of two light flavors at $T^c_0\simeq 170$ MeV. On the other hand, at very low values of $T$ and high values of $\mu_q$ the transition is first order. In addition, from the analysis based on the Hagedorn's limiting temperature~\cite{Hagedorn} at finite $\mu_B$, we know that the critical value for the transition curve to intersect the horizontal axis in the QCD diagram is $\mu_B\simeq m_B$, where $m_B\simeq$ 1 GeV is the typical value of the baryon mass. In one or the other case, since the pion field is a Goldstone mode, the thermal pion mass evaluated at the minima of the potential always vanishes. Then, to fix the coupling constants we use as inputs the values of temperature and quark chemical potential in two extreme points along the transition curve, namely, when the restoration of chiral symmetry is at $\mu_q=0$ (refer as point $A$) and when it is at $T=0$ (refer as point $B$).

At point ($A$), the phase transition is second order, hence the square of the pion thermal mass, evaluated at $v=0$ and $T=T_c^0$, is given by

\begin{equation}
 m_\pi^2(0,T^c_0,\mu_q=0)=-a^2+\Pi(T^c_0,\mu_q=0)=0.
 \label{sigmamassT}
\end{equation}
At point ($B$), the phase transition is first order (degenerates minimal), therefore the minimum we are considering is the one with a vacuum expectation value different from zero, which we call $v_1$. This last condition can be written as

\begin{equation}
 m_\pi^2(v_1,0,\mu_q^c)=\lambda v_1-a^2+\Pi(0,\mu_q^c)=0,
 \label{sigmamassMU}
\end{equation}

In Eq.~(\ref{sigmamassMU}), we notice that a new unknown appears: $v_1$, that is, the value of the non-vanishing minimum. In order to guarantee that the order of the transition in points A and B is consistent with the physical input, we need to add counter-terms $\delta a^2$ and $\delta \lambda$ to the bare constants $a^2$ and $\lambda$, respectively, in the tree level potential 

\begin{align}
	V^{\text{tree}}&=-\frac{a^2}{2}v^2+\frac{\lambda}{4}v^4 \nonumber \\
    &\rightarrow -\frac{(a^2+\delta a^2)}{2}v^2+\frac{(\lambda+\delta \lambda)}{4}v^4.
    \label{newtree}
\end{align}

Therefore the set of conditions necessary to determine $v_1$ and the counter-terms is

\begin{equation}
 \frac{\partial^2 V^{\text{eff}}}{\partial v^2}(v=0,T=T_c,\mu_q=0)=0,
\label{oneminima}
\end{equation}

\begin{align}
 \frac{\partial V^{\text{eff}}}{\partial v}(v=0,T=0,\mu_q=\mu_q^c)=0,&\quad\frac{\partial V^{\text{eff}}}{\partial v}(v=v_1,T=0,\mu_q=\mu_q^c)=0, \nonumber \\
 V^{\text{eff}}(v=0,T=0,\mu_q=\mu_q^c)&=V^{\text{eff}}(v=v_1,T=0,\mu_q=\mu_q^c).
 \label{doubleminima}
\end{align}

The expressions in Eq.~(\ref{oneminima}) indicate that the effective potential is second order when $\mu = 0$ and $T = T_c$ and the three expressions in Eq.~(\ref{doubleminima}) indicate that the effective potential has two degenerated minima at the phase transition and thus that the transition is first order when $T=0$ and $\mu_q=\mu_q^c$. The above set of conditions, Eqs.~(\ref{sigmamassT}),~(\ref{sigmamassMU}) and~(\ref{doubleminima}), represent the five algebraic equations that determine the values of $\lambda$ and $g$. Finally, we can explore the QCD phase diagram.

\section{\label{sec:level8}Results}

Figures~\ref{FPD1}--\ref{FPD2} show the phase diagrams thus obtained. These are computed using the inputs in Table~\ref{tab-1}.  We find that at high (low) temperature and low (high) quark chemical potential the phase transitions are second (first) order. The second order transitions are indicated by the red line and the first order transitions by the blue line. These areas represent the results directly obtained from our analysis.  In all cases, we locate the CEP’s region at low temperatures and high quark chemical potential.
\begin{table}
\centering
\caption{Physical inputs from points A and B and for vaccum bosons masses}
\label{tab-1}       
\begin{tabular}{lll}
\hline
Input & Fig.~\ref{FPD2} & Fig.~\ref{FPD2}  \\\hline
Vaccum $m_\pi$ & 139 MeV & 139 MeV \\
Vaccum $m_\sigma$ & 475 MeV & 475 MeV \\
$T^0_c(\mu_q=0)$ & 170 MeV & 165 MeV \\
$\mu^0_c(T = 0)$ & 340 MeV & 330 MeV \\
$(\lambda,g)$ & (0.87,1.57) & (0.91,1.61) \\\hline
\end{tabular}
\end{table}

\section{\label{sec:level9}Summary}

In this work we have used the linear sigma model with quarks to explore the QCD phase diagram from the point of view of chiral symmetry restoration. We have computed the finite temperature effective potential up to the contribution of the ring diagrams to account for the plasma screening effects. For high quark chemical potential we introduced a boson chemical potentials linked to the high baryon abundanceand.

Our approach was to determine the model's couplings using physical inputs such as the vacuum pion and sigma masses, the LQCD value for the critical temperature at $\mu_q=0$ and the conjectured end point value of $\mu_B$ of the transition line at $T=0$. The set of conditions that determine the couplings enforce the requirement that at high temperature the transition is second order whereas at low temperature is first order.

In order to provide a more robust CEP's location, we need to include the temperature and density modifications to the couplings which has been shown useful to describe the inverse magnetic catalysis phenomenon~\cite{inverse}. For more explicit details, the reader us referred to Ref~\cite{Ayala3}.

\section*{Acknowledgments}

Support for this work has been received in part by UNAM-DGAPA-PAPIIT grant number IN101515 and by Consejo Nacional de Ciencia y Tecnolog\'ia grant number 256494.


\begin{thebibliography}{55}
%
\bibitem{Ayala1} A. Ayala, J. D. Casta\~no-Yepes, J. J. Cobos-Mart\'{\i}nez, S. Hern\'andez-Ortiz, A. J. Mizher, A. Raya, Int. J. Mod. Phys. A {\bf 31}, 1650199 (2016). 

\bibitem{Ayala2} A. Ayala, A. Bashir, J.J. Cobos-Mart\'{\i}nez, S. Hern\'andez-Ortiz, A. Raya, Nucl. Phys. B {\bf 897}, 77-86 (2015).

\bibitem{LeBellac} M. Le Bellac, {\it Thermal Field Theory}  (Cambridge Monographs on Mathematical Physics), Cambridge University Press, 1996. 

\bibitem{DJ} L. Dolan and R. Jackiw, Phys. Rev. D {\bf 9}, 3320-3341 (1974).

\bibitem{chilenos} C. O. Dib, O. R. Espinosa, Nucl. Phys. B {\bf 612}, 492-518 (2001).

\bibitem{latticeTc} T. Bhattacharya {\it et al.}, Phys. Rev. Lett. {\bf 113}, 082001
(2014).

\bibitem{Hagedorn} R. Hagedorn, Nuovo Cim. Suppl. {\bf 3} (1965) 147; Nuovo Cim. A{\bf 56} (1968) 1027; See also K, Fukushima, T. Hatsuda, Rept. Prog. Phys. {\bf 74}, 014001 (2011).

\bibitem{inverse} A. Ayala, L. A. Hern\'andez, M. Loewe, A. Raya, J. C. Rojas, R. Zamora, Phys. Rev. D {\bf 96}, 034007 (2017); A. Ayala, C. A. Dominguez, L. A. Hern\'andez, M. Loewe, A. Raya, J. C. Rojas, C. Villavicencio, Phys. Rev. D {\bf 94}, 054019 (2016); A. Ayala, M. Loewe, R. Zamora, Phys. Rev. D {\bf 91}, 016002 (2015); A. Ayala, M. Loewe, A. J. Mizher, R. Zamora, Phys. Rev. D {\bf 90}, 036001 (2014). 

\bibitem{Ayala3} A. Ayala, S. Hern\'andez-Ortiz and L. A. Hern\'andez, arXiv:1710.09007 (2017).
%


\end{thebibliography}
%
%

\end{document}